\documentclass[aps,prb,reprint,superscriptaddress,showpacs]{revtex4-1}
\usepackage{graphicx}

\begin{document}

\title{High-energy exciton transitions in quasi-two-dimensional cadmium chalcogenide nanoplatelets}

\author{Roman B. Vasiliev}
\email[]{romvas@inorg.chem.msu.ru}
\altaffiliation{Department of Chemistry, Lomonosov Moscow State University, 119991, Moscow, Russia}
\affiliation{Department of Materials Science, Lomonosov Moscow State University, 119991, Moscow, Russia}
\author{Alexander I. Lebedev}
\affiliation{Department of Physics, Lomonosov Moscow State University, 119991, Moscow, Russia}
\author{Elizabeth P. Lazareva}
\affiliation{Department of Materials Science, Lomonosov Moscow State University, 119991, Moscow, Russia}
\author{Natalia N. Shlenskaya}
\affiliation{Department of Materials Science, Lomonosov Moscow State University, 119991, Moscow, Russia}
\author{Vladimir B. Zaytsev}
\affiliation{Department of Physics, Lomonosov Moscow State University, 119991, Moscow, Russia}
\author{Alexei G. Vitukhnovsky}
\altaffiliation{Moscow Institute of Physics and Technology (State University), 141700, Dolgoprudny, Russia}
\affiliation{P.N. Lebedev Physical Institute of the Russian Academy of Sciences, 119991, Moscow, Russia}
\author{Yuanzhao Yao}
\affiliation{Photonic Materials Unit, National Institute for Materials Science, 1-1 Namiki, Tsukuba, Ibaraki 305-0044, Japan}
\author{Kazuaki Sakoda}
\affiliation{Photonic Materials Unit, National Institute for Materials Science, 1-1 Namiki, Tsukuba, Ibaraki 305-0044, Japan}

\date{\today}

\begin{abstract}
Semiconductor nanoparticles of cadmium chalcogenides are known to exhibit
pronounced thickness-dependent $E_0$ series of exciton transitions at the
$\Gamma$ point of the Brillouin zone (BZ). In this work, we report an
experimental evidence for high-energy series of exciton transitions, which
originates from BZ points different from the $\Gamma$ point, in the family
of cadmium chalcogenide quasi-2D nanoplatelets (NPLs). Intensive UV absorption
bands demonstrating a pronounced size effect are observed for CdTe, CdSe, and
CdS NPLs in addition to the $E_0$ exciton bands in the visible region. These
new bands are attributed to transitions analogous to the $E_1$, $E_1+\Delta_1$,
and $E_2$ series observed in bulk crystals. First-principles DFT calculations
of the electronic structure and absorption spectra support this explanation
and show that the main contribution to these optical transitions comes from
$X$ and $M$ points of the 2D BZ, which originate from $L$ and $X$ points of
the 3D BZ. At the same time, the $E_0$ series of transitions at the $\Gamma$
point is well described by the multiband effective-mass model. The observation
of the UV exciton bands reveals tunable optical properties of cadmium
chalcogenide NPLs in UV spectral region, which may be interesting for practical
applications.
\end{abstract}


\maketitle

\section{Introduction}

Colloidal semiconductor nanoparticles have been intensively studied as materials
with promising optical properties tunable with size effects.~\cite{ChemRev.110.389,
NatureMater.13.233,MaterToday.16.312,NaturePhoton.7.13,RussChemRev.80.1139}
In addition to the size, the shape of nanoparticles provides a way to vary their
electronic and optical properties. Substantial progress in the synthesis of
nanoparticles with different shape~\cite{Nature.404.59,ChemMater.18.5722,
NanoLett.10.3770,NatureMater.2.382,NanoLett.5.2164} has been achieved when
growing them in the presence of specific stabilizers. Recently, a considerable
interest has been attracted to cadmium chalcogenide colloidal quasi-2D
nanoplatelets (NPLs) with a thickness of few monolayers.~\cite{JAmChemSoc.130.16504,
NatureMater.10.936} Record narrow absorption and emission bands resulting from
the thickness-dependent exciton transitions make them promising for the
development of new types of light-emitting devices,~\cite{AdvFunctMater.24.295,
NanoLett.15.4611,ChemPhysLett.619.185} lasers,~\cite{NatureNanotechnol.9.891,
NanoLett.14.2772} photodetectors,~\cite{NanoLett.15.1736} photocatalytic
systems,~\cite{JPhysChemLett.6.1099} and as materials with strong electroabsorption
response.~\cite{ACSNano.8.7678}  Atomically flat surface and very homogeneous
thickness of NPLs result in very narrow luminescence bands with a typical width
of 5--10~nm. The giant oscillator strength of the exciton
transitions~\cite{PhysRevLett.59.2337} results in very short luminescence decay
times (hundreds of picoseconds)~\cite{ACSNano.6.6751,NanoLett.13.3321} and
high absorption coefficients for NPLs which are much higher than those for
quantum dots (QDs).~\cite{JPhysChemC.119.20156}

The exciton transitions for cadmium chalcogenides nanoparticles lie in the
visible part of the optical spectrum. However, the ultraviolet (UV) absorption
is of interest too. It is known that cadmium chalcogenide QDs exhibit featureless
absorption at photon energies above 3~eV (Ref.~\onlinecite{JPhysChemB.104.6112})
where the absorption coefficient is independent of their size and proportional
only to the molar fraction of semiconductor. This is why the UV absorption is
often used as a reference for calculating the absorption coefficients for CdSe
and CdTe QDs.~\cite{JPhysChemC.113.6511} In Refs.~\onlinecite{JOptTechnol.78.693,
ACSPhotonics.3.58} this technique was extended to an analysis of the
absorption of core/shell QDs. To the best of our knowledge, the absorption
features of nanoplatelets in the UV spectral region (shorter than 380--400~nm)
have not been studied yet.

In this work, we report an observation of new high-energy exciton transition
series in the family of cadmium chalcogenide NPLs. We analyzed the absorption
and photoluminescence excitation spectra of CdS, CdSe, and CdTe NPLs of different
thickness capped with various ligands in the UV region and revealed pronounced
and intensive thickness-dependent exciton bands in addition to exciton bands
in the visible region. First-principles DFT calculations of the electronic
structure and absorption spectra of NPLs shows that these high-energy bands
result from transitions at the $X$ and $M$ points of the 2D Brillouin zone (BZ)
of NPLs, which originate from $L$ and $X$ points of the BZ of bulk crystals.
At the same time, the series of transitions at the $\Gamma$ point in the visible
region is well described by the multiband effective-mass model.

\section{Experiment}

\subsection{Nanoplatelet synthesis}

Colloidal technique with cadmium acetate as an agent promoting the formation of
2D nanoparticles was used to synthesize nanoplatelets. The CdTe NPLs (the
populations with the main exciton emission bands at 430, 500, 556, 600, and
634~nm that are further referred to as CdTe430, CdTe500, etc.), CdSe NPLs
(CdSe396, CdSe463, CdSe512, and CdSe550) and CdS NPLs (CdS380) were synthesized.
The synthesis of CdTe NPLs (CdTe430, CdTe500, and CdTe556 populations) was
performed according to the method proposed in Ref.~\onlinecite{ChemMater.25.2455}.
A mixture of thicker CdTe600 and CdTe634 populations was obtained by slow heating
of the reaction mixture to 270~$^\circ$C. The synthesis of CdSe NPLs (CdSe512
and CdSe550 populations) was carried out following the protocols given in
Ref.~\onlinecite{JAmChemSoc.133.3070}.  The modified technique adapted from
Refs.~\onlinecite{JAmChemSoc.133.3070,NanoRes.5.337} was used for the
synthesis of CdSe NPLs (CdSe396 and CdSe463 populations) and CdS NPLs (the
CdS380 population). Briefly, a mixture containing 0.13~g of cadmium acetate
dihydrate, 0.08~ml of oleic acid, and 10~ml of octadecene was heated to
230~$^\circ$C (CdSe463 and CdS380) or 130~$^\circ$C (CdSe396) under
argon flow. After that, 100~$\mu$l of 1M solution of selenium or sulfur in
trioctylphosphine was injected and the growth of the NPLs was continued for 30~min
(CdSe463 and CdS380) or 6~hours (CdSe396). Then the reaction mixture was cooled
down to the room temperature. During the cooling process, 1~ml of oleic acid was
injected and nanoplatelets were precipitated by acetone. After 2--3 cycles of
precipitation and redispersion, the solutions of the NPLs in hexane with
minimum impurity content were obtained. Ligand exchange with the thioglycolic
acid (TGA) was applied to synthesize a CdSe463TGA sample using phase-transfer
method similar to Ref.~\onlinecite{JAmChemSoc.134.18585}.

\subsection{Measurement and calculations}

Micrographs of nanoplatelet ensembles were recorded on LEO912 AB OMEGA transmission
electron microscope. The accelerating voltage was 15~kV.

Absorption spectra were recorded using the Varian Cary 50 spectrophotometer on
nanoplatelets dispersed in spectroscopic grade solvents (hexane or methanol) in
the spectral range of 200--800~nm. The spectra were analyzed with the PeakFit
software and were fitted with Lorentz profile bands. In some cases, the background
caused by light scattering was subtracted. The photoluminescence excitation
spectra were collected using the Perkin-Elmer LS 55 fluorescence spectrometer
by monitoring the intensity of the lowest-energy exciton luminescence band.

First-principles calculations of the band structure and absorption spectra of
CdSe NPLs were carried out within the density-functional theory using the
\texttt{ABINIT} software. In order to correctly account for spin-obit coupling,
the LDA PAW pseudopotentials taken from Ref.~\onlinecite{ComputMaterSci.81.446}
were used. The cut-off energy was 30~Ha, the integration over the Brillouin zone
was performed on a 8$\times$8$\times$2 Monkhorst-Pach mesh.

The absorption spectra were calculated using the \texttt{OPTIC} program of the
\texttt{ABINIT} package by summing the contributions of all band-to-band optical
transitions into the imaginary part of the dielectric function on a dense
${\bf k}$-point mesh in the Brillouin zone. In order to correctly take into
account all possible electronic transitions, up to 30~empty conduction bands
were used in these calculations. The 26$\times$26$\times$26 ${\bf k}$-point
mesh was used for bulk CdSe and 26$\times$26$\times$1 or denser ${\bf k}$-point
meshes were used for NLPs. During the analysis, small changes in the \texttt{OPTIC}
code were made in order to isolate the contributions from given points of the BZ
to the absorption spectrum.

To calculate the thickness dependence of the exciton transition energies
resulting from the optical transitions at the $\Gamma$ point, we used the
multiband effective-mass calculation method which was previously used by
Ithurria {\it et al.} for the study of CdSe, CdS, and CdTe
NPLs.~\cite{NatureMater.10.936}

\section{Results}

\subsection{TEM studies}

\begin{figure}
\includegraphics[width=0.49\linewidth]{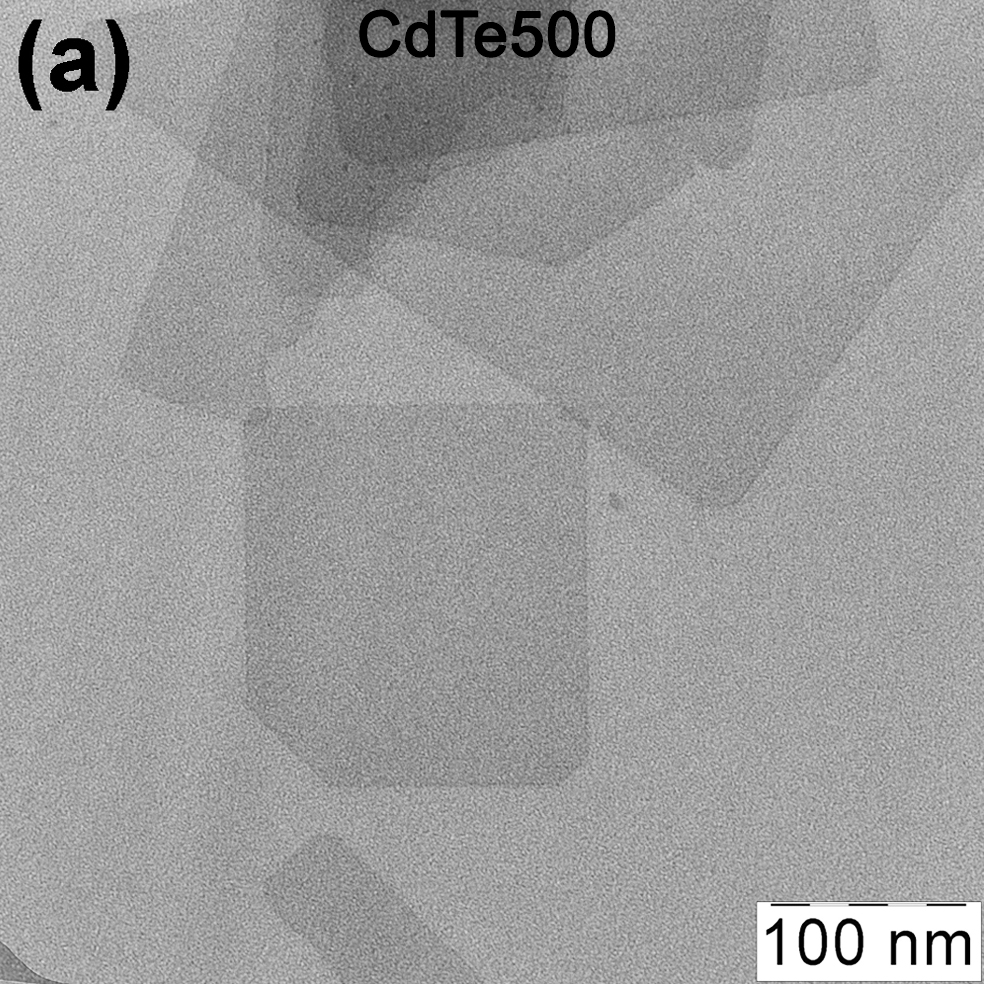}
\includegraphics[width=0.49\linewidth]{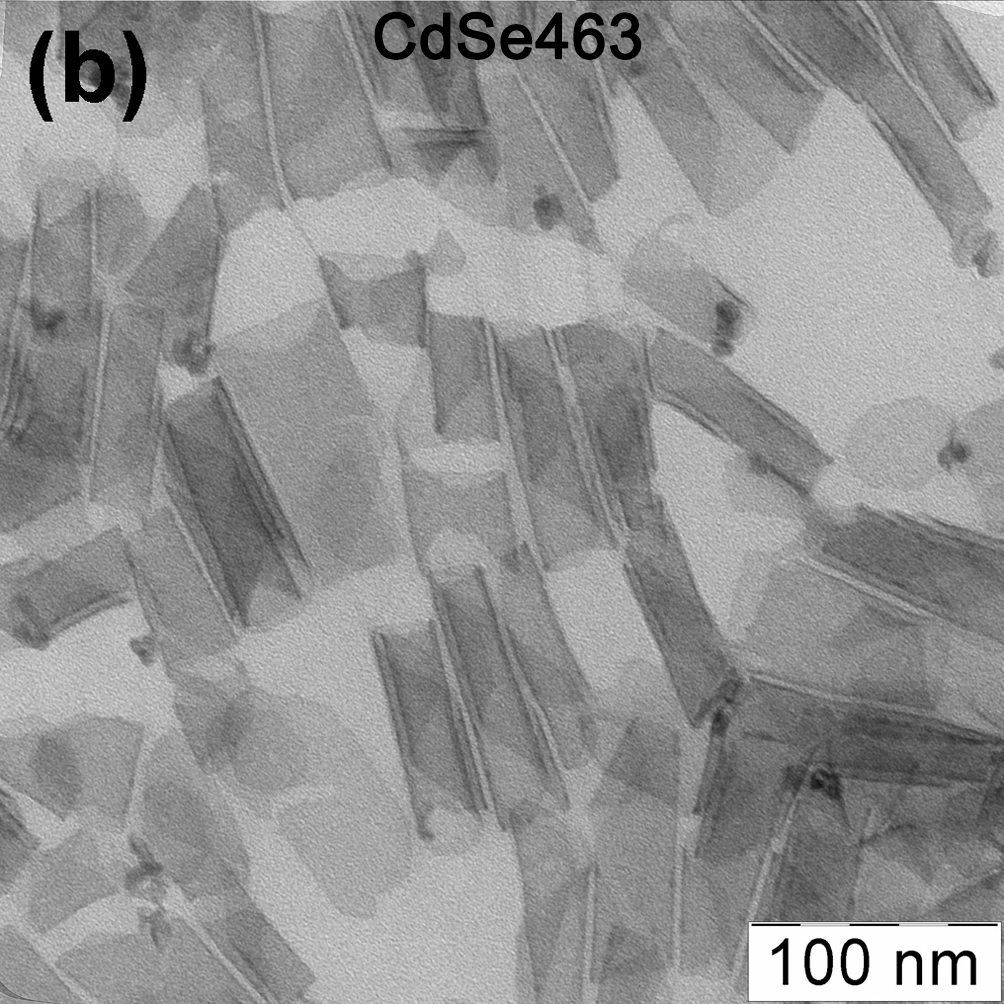}
\includegraphics[width=0.49\linewidth]{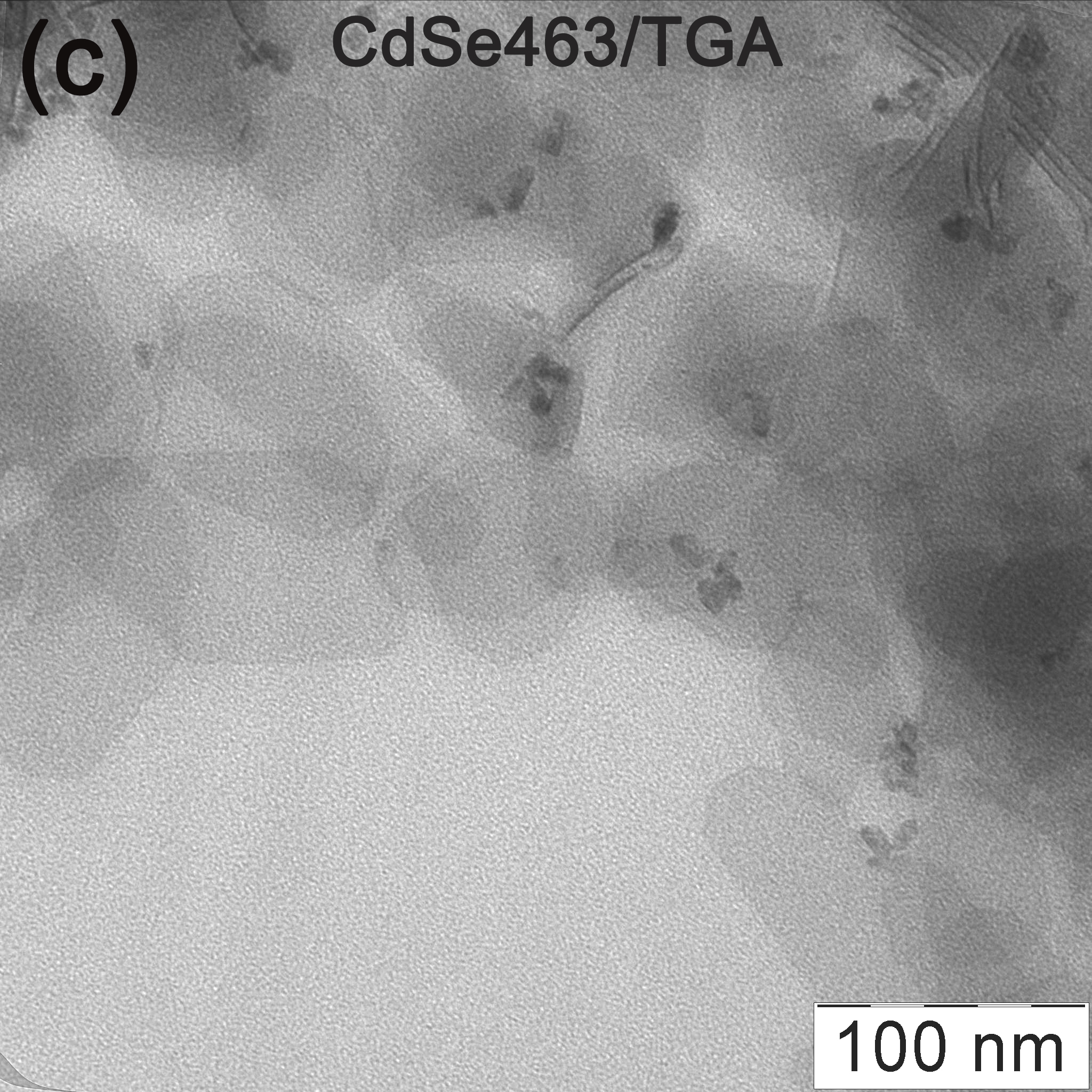}
\includegraphics[width=0.49\linewidth]{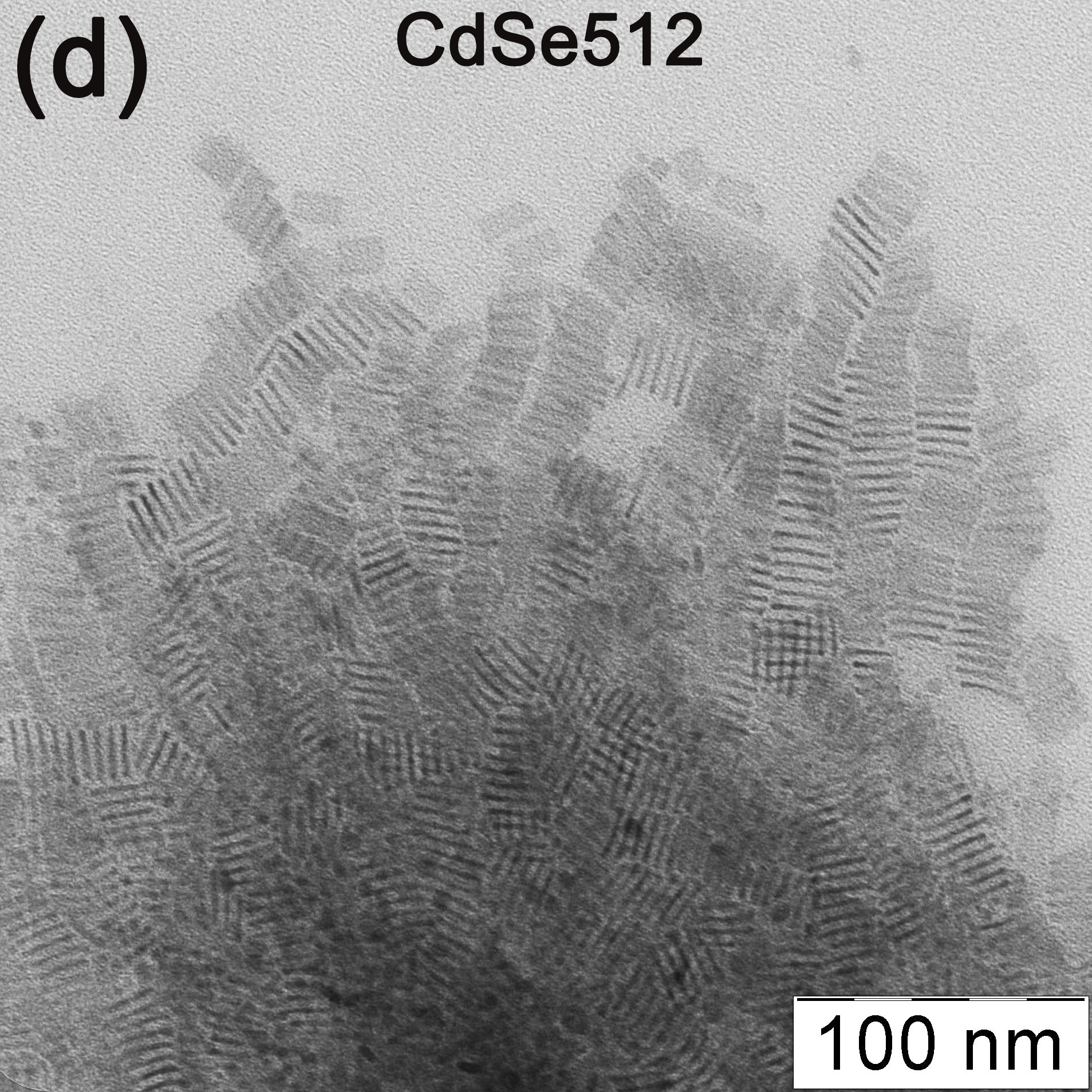}
\includegraphics{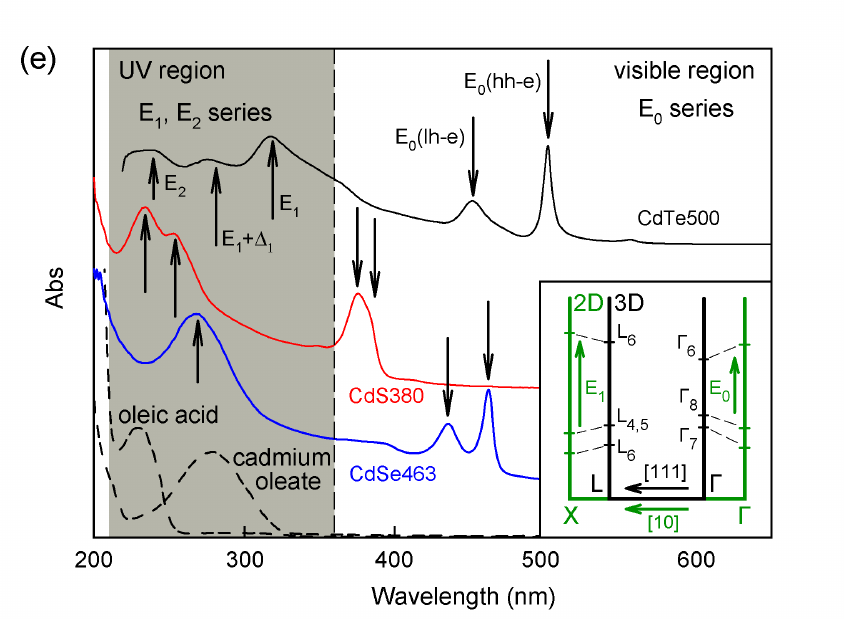}
\caption{\label{fig1}Typical TEM images of (a)~CdTe500, (b)~CdSe463,
(c)~CdSe463TGA, and (d)~CdSe512 NPLs. (e)~Absorption spectrum of CdTe500,
CdS380, and CdSe463 NPLs. Absorption spectra of cadmium oleate and oleic
acid are shown by dashed lines. Spectral positions of exciton bands are
indicated with arrows. Inset: Schematic band diagram showing the correspondence
between the exciton transitions at $L$ and $\Gamma$ points of the Brillouin
zone in bulk samples and at $X$ and $\Gamma$ points of the Brillouin zone
for quasi-2D NPLs.}
\end{figure}

TEM studies of obtained NPLs showed that their lateral size was 100--200~nm
except for CdSe512 and CdSe550 populations whose size was about 10~nm
[Fig.~1(a--d)]. CdSe NPLs with large lateral size were observed to roll-up
into nanoscrolls, however, after TGA treatment they unfolded to flat
nanoplatelets, in agreement with the literature data.~\cite{ChemMater.25.639}
According to X-ray and electron diffraction measurements, all the NPLs had the
zinc-blende crystal structure.

\subsection{Optical absorption and photoluminescence studies}

Typical absorption spectra of CdTe, CdSe, and CdS NPLs are shown in Fig.~1(e).
Two well-defined narrow absorption bands for each sample appear in the visible
region. These bands correspond to the transitions from the light-hole and
heavy-hole valence sub-bands to the conduction band and are denoted as
$E_0(lh$--$e)$ and $E_0(hh$--$e)$, respectively. The spectral positions of these
bands are consistent with the literature data.~\cite{JAmChemSoc.130.16504,
NatureMater.10.936,ChemMater.25.2455,NanoRes.5.337}  In the UV
region (200--400~nm), however, unexpectedly non-monotonic spectra with pronounced
and intensive absorption bands was revealed, in contrast with the featureless
absorption spectra typical of spherical QDs. The observed fine structure was
different for various NPLs. For CdTe NPLs, three absorption bands with the width
of 50--70~nm can be distinguished. These bands are wider than the $E_0(lh$--$e)$
and $E_0(hh$--$e)$ bands, whose widths are 6--8 and 15--20~nm, respectively. It
should be noted that the shape of the UV bands is close to the Lorentz profile,
which indicates the absence of inhomogeneous broadening. The UV absorption
spectra of CdSe and CdS NPLs reveal a similar behavior. A single maximum was
observed for CdSe463 and CdSe396 NPL populations, whereas two maxima were
registered for CdS NPLs.

\begin{figure}
\includegraphics{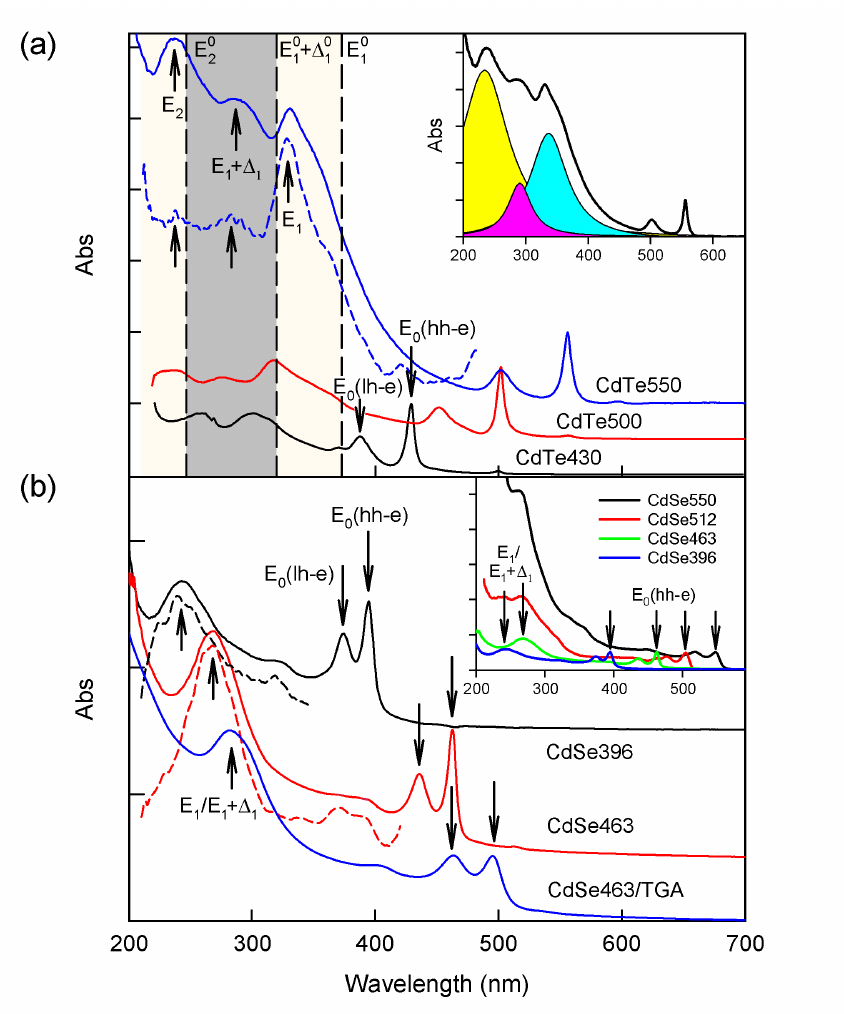}
\includegraphics{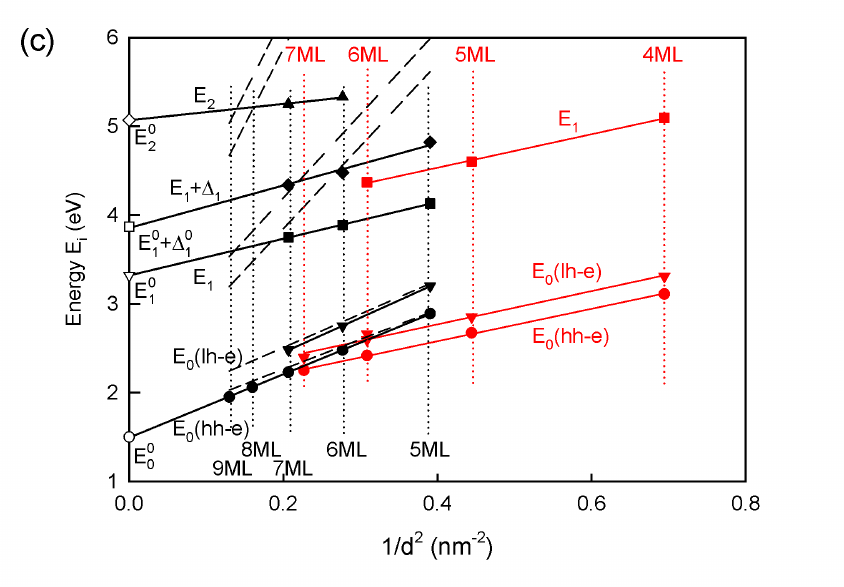}
\caption{\label{fig2}(a)~Absorption (solid lines) and PL excitation (dashed lines)
spectra of CdTe NPLs with different thickness. Dashed vertical lines highlight
the positions of $E_1^0$, $E_1^0+\Delta_1^0$, and $E_2^0$ transitions for bulk
CdTe. The spectra are offset for clarity. The inset presents the CdTe550 absorption
spectrum fitted with three Lorentz profiles. (b)~Absorption and PL excitation
spectra of CdSe NPLs with different thickness. Spectral positions of the exciton
bands are indicated with arrows. Inset: absorption spectra of large-area CdSe396
and CdSe463 NPLs and small-area CdSe512 and CdSe550 populations. (c)~Dependence
of the exciton transition energies $E_i$ on the reciprocal square of thickness $d$
for CdTe (black color) and CdSe (red color) NPLs. The $E_i^0$ values for CdTe are
shown by open symbols. Solid lines are linear approximations of experimental data,
dashed lines show the exciton energies predicted by the multiband effective-mass
model.}
\end{figure}

The absorption spectra of CdTe NPLs with different thickness are shown in
Fig.~2(a). For all populations, three bands in UV region are observed. These
bands shift to longer wavelengths with increasing thickness.

The presence of the absorption bands is supported by photoluminescence (PL)
excitation spectroscopy. The PL excitation spectrum for CdTe550 NPL shown by
dashed line in Fig.~2(a) clearly demonstrates the features coinciding well with
those in the absorption spectrum. The PL excitation spectra and the
thickness-dependent absorption spectra exclude a possible interpretation of
the UV absorption bands as resulting from the traces of free oleic acid and
cadmium oleate [Fig.~1(e)].

In the case of CdSe NPLs, a pronounced size effect is observed for CdSe396 and
CdSe463 populations [Fig.~2(b)]. A single UV band shifts to higher energies with
decreasing thickness. The PL excitation spectra also show a pronounced peak in
the UV region which coincides with the absorption bands in both the samples.
Additional bands at 370 and 392~nm (CdSe463) and at 319~nm (CdSe396) observed
in PL excitation spectra can be attributed to the $E_0(2s)$ series transitions
analogous to those observed in epitaxial quantum wells~\cite{PhysRevB.51.14395}
or an exciton formed by an electron and a hole from spin-orbit split
band.~\cite{ChemMater.25.1190}
However, this behavior does not apply to CdSe512 and CdSe550 populations whose
high-energy peaks lie at higher energies as compared to their expected positions
[Fig.~2(b)]. This could be a result of significantly smaller lateral size of
these NPLs which is comparable to the Bohr radius of exciton in CdSe (7~nm)
and so the expected 2D behavior is violated. To get additional information on
the origin of the UV bands, we performed ligand exchange with thioglycolic
acid to synthesize a CdSe463TGA sample covered with ligands other than the
oleic acid. In this sample, the single UV band is retained but is shifted to
longer wavelength [Fig.~2(b)] because of an increase in the NPL thickness
resulted from the sulfide layer formation.

\subsection{First-principles calculations of the band structure and absorption spectra}

To calculate the band structure of NPLs, periodic structures consisting of slabs
of (001)-oriented NPLs with the zinc-blende structure separated by vacuum layers
were used. The nanoplatelets had a thickness n from one to six monolayers (ML)
and were terminated by Cd atoms on both sides. The vacuum layer of 20~{\AA}
between the slabs was found to be sufficient to neglect the interaction between
the NPL and its images. To make the nanoplatelets insulating, two terminating
F atoms were added near the Cd atoms on both surfaces of the NPL to compensate
the charge of $s^2$ electrons produced by an extra Cd plane. The energies of
the on-top, hollow, and four different bridge configurations of F atoms on the
surfaces of NPLs were compared; the bridge position with F atoms located near
the expected Se sites of the zinc-blende structure was found to be the
ground-state structure with the $P{\bar 4}m2$ space group.

\begin{figure*}
\includegraphics{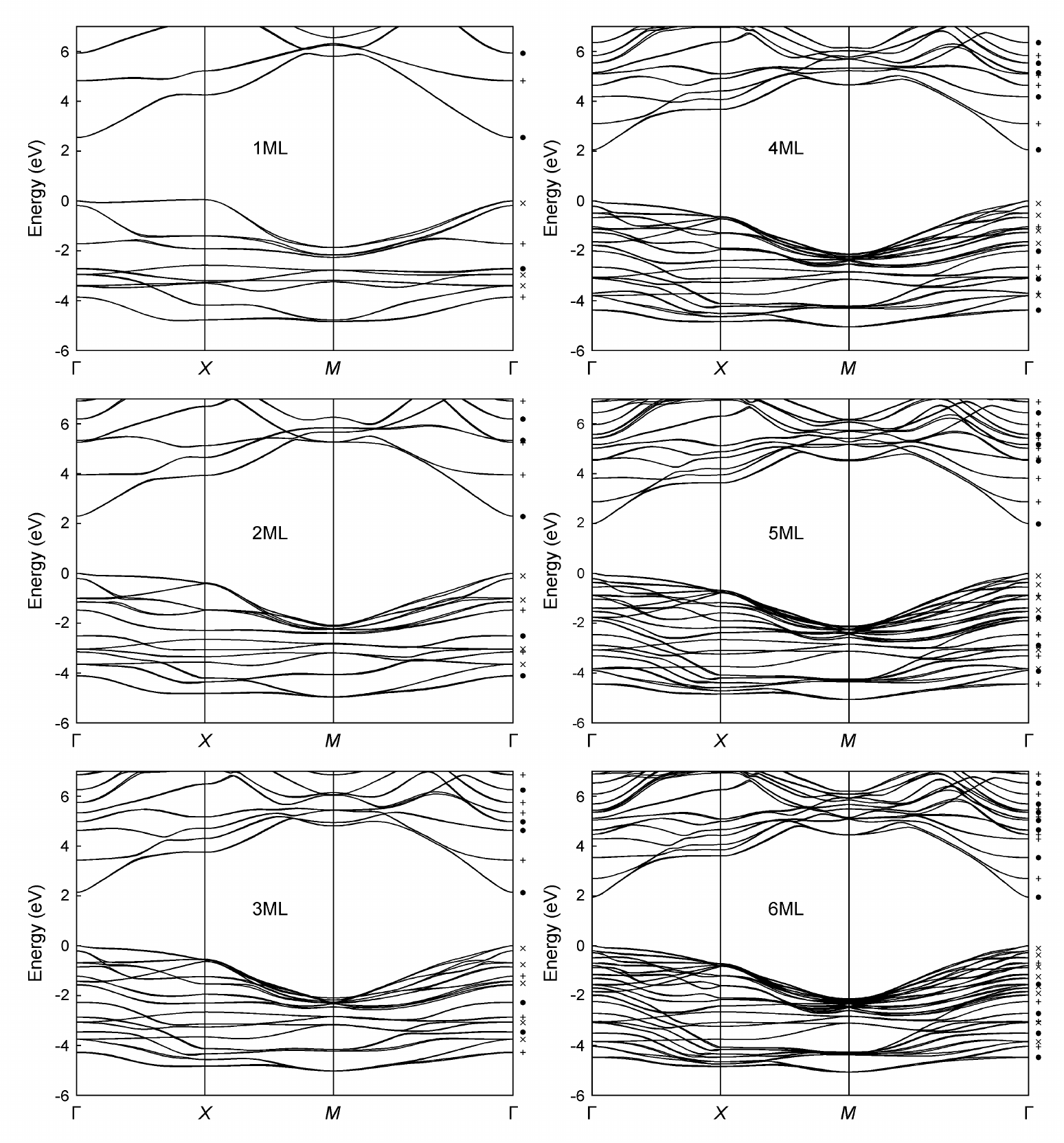}
\caption{\label{fig3}Evolution of the band structure of CdSe NPLs with increasing
their thickness form one to six monolayers. The symbols at the right side of each
panel denote the symmetry of wave functions at the $\Gamma$ point:
($\bullet$) $A_1$, (+) $B_2$, and ($\times$) $E$.}
\end{figure*}

At first, the geometry of NPLs (the in-plane lattice parameter and atomic
positions) was carefully optimized until the residual forces acting on the
atoms decreased to below $5 \cdot 10^{-6}$~Ha/Bohr. The in-plane lattice
parameter of NPLs was found to increase from 3.9513~{\AA} for 1ML to
4.1731~{\AA} for 6ML toward the calculated interatomic Cd--Cd distance of
4.2544~{\AA} for bulk CdSe. The band structure of all NPLs was then calculated
(Fig.~3). The scissors correction of 1.495~eV estimated from the band structure
calculations for bulk zinc-blende CdSe was applied to the energies of all
conduction bands of NPLs (the applicability of the scissors correction to
CdSe was recently approved by $GW$ calculations~\cite{PhysRevB.50.10780}).

An analysis of the band structures shows that the valence band of
Cd$_{n+1}$Se$_n$F$_2$ NPLs is composed of $3(n+2)$ bands with the $A_1$, $B_2$,
and $E$ symmetry at the $\Gamma$ point originating mainly from F $2p$ and Se
$4p$ atomic states. Note a large spin-orbit splitting for $E$ bands originating
from Se states (upper part of the valence band) and negligibly small splitting
for $E$ bands originating from F states (lower part of the valence band).

\begin{figure}
\includegraphics{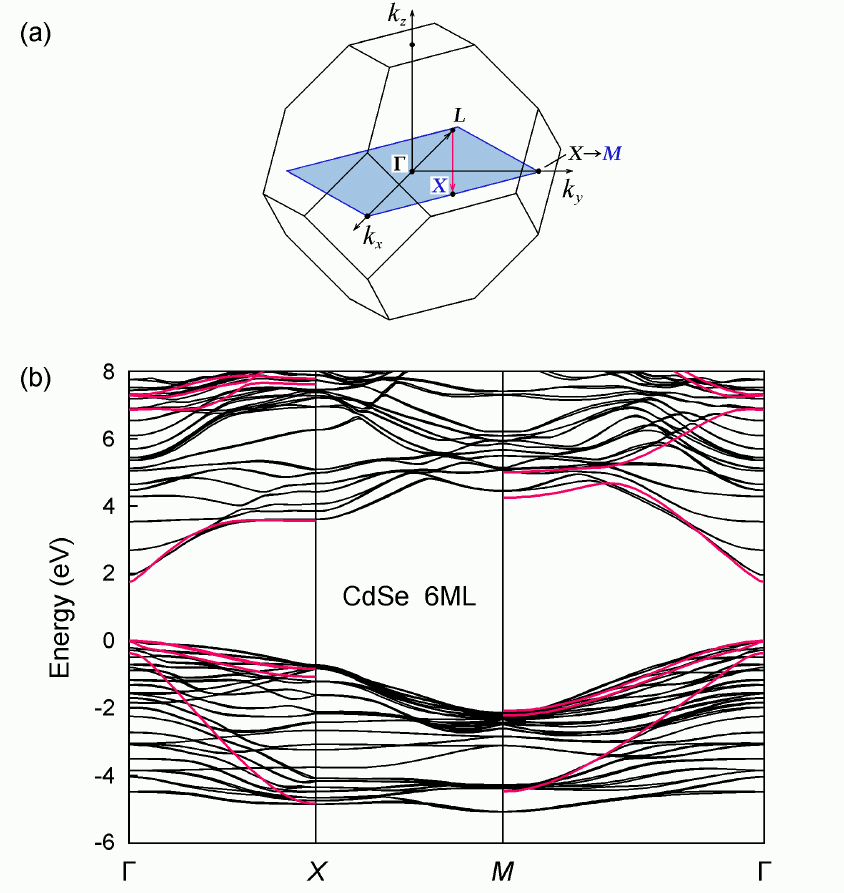}
\includegraphics{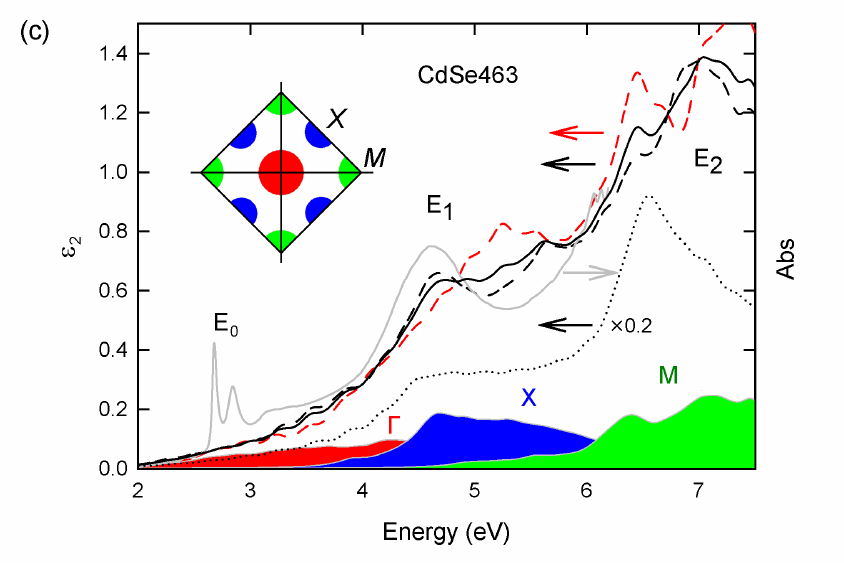}
\caption{\label{fig4}(a)~The Brillouin zone for 3D zinc-blende structure and its
2D projection. (b)~Comparison of the band structures of 6ML CdSe NPL (black lines)
and bulk CdSe (red lines). (c)~Comparison of the experimental and modeled
absorption spectra for CdSe463 NPLs (see text for explanation).}
\end{figure}

The structure of the conduction band of NPLs appeared to be different from
that expected in simple theories that predict size-effect-induced shifts of
dispersion curves for sub-bands. In our case, we see a number of the
lower-lying bands with the $A_1$ and $B_2$ symmetry at the $\Gamma$ point
originating mainly from, respectively, $5s$ and $5p_z$ empty atomic orbitals
of Cd and Se. The higher-lying conduction bands have predominantly the $E$
symmetry and originate from Cd and Se $5p_{x,y}$ states. An inspection of
Fig.~3 and a comparison between the band structures of 6ML NPLs and bulk
CdSe [Fig.~4(b)] shows that the number of bands below the $E$ bands can be
as large as 10--14. It looks like one of three $5p$ orbitals of Se, $p_z$,
is strongly split off and form these extra bands while the other $5p_x$
and $5p_y$ orbitals form the higher-lying bands with wave functions of the
$E$ symmetry. As a result of a complex interaction between $s$ and $p_z$
atomic orbitals, a number of subbands have properties that are very different
from those expected in simple theories. For example, these bands have very
different effective masses (which is unusual for sub-bands). The comparison
of the remaining parts of the band structures shows that there is a
qualitative agreement between them: we can see the formation of dense
``nodes'' in the valence band along the dispersion curves of bulk CdSe when
the number of monolayers is increased [see Fig.~3 and Fig.~4(b)].

\section{Discussion}

The $E_0(lh$--$e)$ and $E_0(hh$--$e)$ bands in the visible region can be attributed
to the $E_0$ exciton series at the $\Gamma$ point of the BZ [Fig.~1(c), inset]
and are related to the fundamental absorption edge. In the UV region, the behavior
of the absorption bands is different. Although for different chalcogenides
different fine structure of these bands was observed, in all cases the absorption
fine structure in NPLs looks very similar to that in bulk
crystals.~\cite{JApplPhys.78.4681,JApplPhys.74.3435,PhysRevB.49.7262,JApplPhys.78.1183}
Three bands observed for CdTe NPLs in the UV region may be associated with $E_1$,
$E_1+\Delta_1$, and $E_2$ transitions found in bulk CdTe.~\cite{JApplPhys.78.4681}
The single UV band observed for CdSe NPLs can be explained by an overlapped $E_1$
and $E_1+\Delta_1$ transitions resulting from weaker spin-orbital coupling. This
agrees with the optical spectra for bulk zinc-blende CdSe in the UV region,
in which a pronounced $E_1$ peak and a shoulder for $E_1+\Delta_1$ transition
were observed.~\cite{JApplPhys.74.3435,PhysRevB.49.7262}  Two UV bands observed
for CdS NPLs are in agreement with two maxima observed for
CdSe~\cite{PhysRevB.49.7262} (however, it should be noted that the data
in Ref.~\onlinecite{JApplPhys.78.1183} are obtained for wurtzite CdS). In all
cadmium chalcogenide NPLs, the UV bands are shifted towards shorter wavelengths
as compared to the data for bulk compounds. In Fig.~2(c), we plotted their
energies $E_i$ as a function of the reciprocal square of the thickness $1/d^2$.
The NPL thickness values were taken from the literature.~\cite{NatureMater.10.936,
ChemMater.25.2455}  The linear dependence of $E_i$ on $1/d^2$ for all exciton
series confirms the size effect. The asymptotic energies in the $1/d^2 \to 0$
limit are close to the energies $E_i^0$ of the corresponding transitions in
bulk compounds.~\cite{JApplPhys.78.4681,JApplPhys.74.3435}  For UV bands, the
thickness dependence of the shifts is weaker than that for the $E_0$ series;
this may indicate a larger effective mass of charge carriers.

The hypothesis about the relation between the UV absorption bands and optical
$E_1$ and $E_2$ transitions at the boundary of the BZ needs a detailed analysis.
For bulk cadmium chalcogenides with zinc-blende structure, these high-energy
transitions are known to occur at the $L$ point ($E_1$ and $E_1+\Delta_1$
transitions) and the $X$ point ($E_2$ transition) of the 3D BZ. In the case of
quasi-2D nanoplatelets, the 3D BZ transforms to its 2D projection [Fig.~4(a)]:
the $X$ point becomes the $M$ point at corner of square, and the $L$ point (which
does not exist in 2D BZ) is projected to the $X$ point at the middle of its
side. The comparison of the band structures for a typical F-terminated CdSe
NPL with a thickness of 6~ML and bulk CdSe [Fig.~4(b)] shows a good agreement
between the dispersion curves at the band edges. This means that the energies
of the band extrema at the boundary of the 2D BZ are close to those in the band
structure of bulk CdSe. This fact supports our interpretation of the UV absorption
bands as resulting from the optical transitions at the $X$ and $M$ points of the
2D BZ.

In order to give further evidences for our interpretation, we performed
first-principles modeling of the absorption spectrum for CdSe463 NPL. The
calculated absorption spectra for the in-plane and out-of-plane light
polarizations (red and black dashed lines), the absorption spectrum averaged
over polarizations (solid black line), and partial contributions into this
spectrum from different points of 2D Brillouin zone (color shaded areas; the
BZ regions used for integration are shown by shaded areas on the inset) are
compared with the experimental spectrum (grey line) and the calculated
absorption spectrum for bulk CdSe (dotted line) in Fig.~4(c). It is evident
that the main contributions to the $E_1$ and $E_2$ bands originate from $X$
and $M$ points of the BZ, respectively, not from the $\Gamma$ point. This
explains weaker thickness-dependent shifts observed for these transitions.
The agreement between the modeled and experimental spectra in the $E_1$ region
is good. The position of the $E_2$ band lies outside of our experimentally
available wavelength range. The excitonic features in the low-energy $E_0$
part of the spectrum are not reproduced because our calculations did not take
into account the excitonic effects. It should be noted that the appearance of
distinct $E_1$ and $E_2$ exciton peaks in the absorption spectra of NPLs is due
to the specific features of their electronic structure that results in
step-like or divergent van Hove singularities.~\cite{PhysRevB.51.14395}

\begin{table}
\caption{The band structure parameters for CdTe and CdSe used in the
multiband model.}
\label{table1}
\begin{ruledtabular}
\begin{tabular}{ccc}
Parameter & CdTe & CdSe \\
\hline
Band gap $E_g$ (eV)    & 1.49 & 1.66 \\
Kane energy $E_p$ (eV) & 21   & 16.5 \\
Spin-orbit splitting $\Delta$ (eV) & 0.6 & 0.39 \\
$\alpha$   & $-$0.13 & $-$1.54 \\
$\gamma_1$ & 0.99    & $-$0.18 \\
$\gamma_2$ & $-$0.52 & $-$0.65 \\
$m_e$ ($m_0$) & 0.11 & 0.183 \\
$m_{hh}$ ($m_0$) & 0.69 & 0.89 \\
Dielectric constant $\epsilon$ & 10.4 & 9.6 \\
Lattice constant (nm) & 0.648 & 0.608 \\
\end{tabular}
\end{ruledtabular}
\end{table}

The contribution of the exciton transitions at the $\Gamma$ point was analyzed
within the eight-band Pidgeon--Brown model. In this model, the bulk band
dispersion is characterized in terms of the ${\bf k \cdot p}$ band parameters,
which include the bulk band gap energy $E_g$, the Kane energy $E_p$, the
valence band spin-orbit splitting $\Delta$, the remote band contribution to the
electron effective mass $\alpha$, and the valence band Luttinger--Kohn parameters
$\gamma_1$ and $\gamma_2$. The parameters used for CdTe and CdSe NPLs were taken
from Refs.~\onlinecite{NatureMater.10.936,Adachi2009} and are summarized in
Table~\ref{table1}. The effective-mass approximation was applied with an
assumption of an infinite potential-barrier height and the energies of low-energy
$E_0(lh$--$e)$ and $E_0(hh$--$e)$ transitions were calculated. We assumed that
the in-plane wave vector ${\bf k}_\parallel = 0$ and the out-of-plane wave vector
component ${\bf k}_\perp = N\pi/d$, where $N$ is the confinement
quantum number and $d$ is the thickness of the NPL. The thickness dependence
of exciton transition energies was found for $N = 1, 2, 3$. The calculated
energies for CdTe NPLs are shown in Fig.~2(c) by dashed lines. Lowest-energy
transitions are in good agreement with the experimental data. A similar
behavior was observed for CdSe NPLs. However, the thickness dependence of
the higher-order transitions deviates from the experimentally observed one,
which possibly indicates that higher-order transitions are weaker than the
transitions at the boundary of BZ.

\section{Conclusions}

In summary, we have revealed that cadmium chalcogenide NPLs exhibit distinct
and intensive exciton bands with a pronounced size effect not only in the
visible region, but also in the UV region. We have shown that these
high-energy bands are associated with the $E_1$, $E_1+\Delta_1$, and $E_2$
exciton transitions occurring at the $X$ and $M$ points of the 2D BZ of
NPL, which originates from $L$ and $X$ points of the BZ of bulk crystals.
This behavior is general for all 2D NPLs and contrasts with the behavior
of spherical QDs made of cadmium chalcogenides. Thickness-dependent shift
of the absorption bands in the NPLs reveals tunable optical properties of
them not only in the visible, but in the UV spectral region too. The specific
features of the UV absorption spectra resulting from step-like or divergent
van Hove singularities typical of quasi-2D NPLs may attract a fundamental
interest to an analysis of the electronic transitions in other 2D semiconductor
systems deep into the Brillouin zone. The UV features of the NPL absorption
spectra may be interesting from a practical perspective for the development
of UV photodetectors and other optoelectronic devices.

\begin{acknowledgments}
This work was supported by the Russian Foundation for Basic Research (grants
No.~16-03-00704, 16-29-11694, 15-02-05856, and 16-29-11805).
\end{acknowledgments}

%

\end{document}